\documentclass[12pt,preprint]{aastex}

\def\geqsim{\lower.73ex\hbox{$\sim$}\llap{\raise.4ex\hbox{$>$}}$\,$}
\def\leqsim{\lower.73ex\hbox{$\sim$}\llap{\raise.4ex\hbox{$<$}}$\,$}

\def\Mpchseventy {~h_{70}^{-1}~{\rm Mpc}}
\def\kpch {~h^{-1}~{\rm kpc}}
\def\kpchseventy {~h_{70}^{-1}~{\rm kpc}}

\slugcomment{ApJ, accepted 24 July 2008}
\shorttitle{AGN Environments}
\shortauthors{Strand et al.}

\begin{document}

\title{AGN Environments in the Sloan Digital Sky Survey I:\\ Dependence on Type, Redshift, and Luminosity}
\author{Natalie E. Strand}
\affil{Department of Physics, University of Illinois at Urbana-Champaign}
\affil{Urbana, IL 61801}
\email{nstrand@illinois.edu}
\and
\author{Robert J. Brunner, Adam D. Myers}
\affil{Department of Astronomy, University of Illinois at Urbana-Champaign}
\affil{Urbana, IL 61801}

\begin{abstract}

We explore how the local environment is related to the redshift, type, and luminosity of active galactic nuclei (AGN).  Recent simulations and observations are converging on the view that the extreme luminosity of quasars is fueled in major mergers of gas-rich galaxies.  In such a picture, quasars are expected to be located in regions with a higher density of galaxies on small scales where mergers are more likely to take place.  However, in this picture, the activity observed in low-luminosity AGN is due to secular processes that are less dependent on the local galaxy density.  
To test this hypothesis, we compare the local photometric galaxy density on kiloparsec scales around spectroscopic Type I and Type II quasars to the local density around lower luminosity spectroscopic Type I and Type II AGN.  To minimize projection effects and evolution in the photometric galaxy sample we use to characterize AGN environments, we place our random control sample at the same redshift as our AGN and impose a narrow redshift window around both the AGN and control targets.  
We find that higher luminosity AGN have more overdense environments compared to lower luminosity AGN on all scales out to our $2\Mpchseventy$ limit.  
Additionally, in the range $0.3\leqslant z\leqslant 0.6$, Type II quasars have similarly overdense environments to those of bright Type I quasars on all scales out to our $2\Mpchseventy$ limit, while the environment of dimmer Type I quasars appears to be less overdense than the environment of Type II quasars.  
We see increased overdensity for Type II AGN compared to Type I AGN on scales out to our limit of $2\Mpchseventy$ in overlapping redshift ranges.  
We also detect marginal evidence for evolution in the number of galaxies within $2\Mpchseventy$ of a quasar with redshift.  

\end{abstract}


\section{INTRODUCTION}\label{introsection}

     The local environments of quasars provide valuable insight into the formation history and evolution of AGN \citep[e.g.,][]{Ellingson}.  Quasar environments were first studied by \citet{Bahcall}, who used a sample of five quasars to show that quasars are associated with galaxy clusters.  \citet{YeeGreen} found that quasars reside in regions with higher galaxy density, and more recent work has confirmed that quasars are found in regions of galaxy groups or clusters of poor to moderate richness \citep{BahcallChokshi, Fisher, McLureDunlop, Wold, Coldwell2002, Barr}.  Although studies of several X-ray- and radio-selected samples have found evidence for a relationship between environment and AGN activity \citep[e.g.,][]{Wurtz, Best, Sochting}, the Sloan Digital Sky Survey (SDSS) is the first survey to allow meaningful studies of quasar environments, because it samples large numbers of both quasars and galaxies at $z \lesssim 0.4$.  Using SDSS data, \citet{Serber} concluded that the density of photometric galaxies around quasars increases with decreasing angular scale, but is independent of redshift for $z \leqslant 0.4$.  They also provided evidence for a higher density of galaxies around more luminous quasars at scales less than 100$\kpch$, while at larger angular scales, the density appears to be largely independent of luminosity \citep[c.f.][]{PorcianiNorberg, daAngela, Myers07a}.  

     The results of \citet{Serber} agree with other studies showing enhanced clustering of quasars on small scales.  \citet{Djorgovski} first linked the excess of quasar clustering on small scales to galaxy interactions.  Studies of small-scale clustering of quasars (e.g., binary and triplet quasars) also support the hypothesis that there is excess quasar clustering on scales of $\lesssim 100$$\kpch$ \citep{Kochanek, Mortlock, Hennawi, Djorgovski2007, Myers07a}.
    
    An excess of quasar pairs on small scales naturally follows from a merger origin for quasar activity, whether these pairs simply trace biased groups where mergers are likely to occur \citep{Hopkins2007} or are being excited in merging galaxies \citep{Djorgovski, Myers07b}.  \citet{Hopkins2006} have developed a unified, merger-driven framework that naturally predicts that quasar environments should be highly biased \citep{Hopkins2007}.  These simulations show that major mergers between gas-rich galaxies are the likely mechanisms to trigger bright quasar activity, and that this activity is a phase in the evolution of massive spheroidal galaxies \citep{Hopkins2005b, Hopkins2007}.  In contrast, secular mechanisms may fuel the activity in most low-luminosity AGN, implying that the small-scale environments of these objects should have a smaller bias \citep{HopkinsHernquist, Hopkins2007}.  Therefore, we would expect that objects driven by major mergers will have biased environments on small scales, whereas objects fueled by secular means will reside in less rich environments.  Such a simplification hides many subtleties, however, as secular mechanisms such as harassment can probably only occur in slightly overdense environments.  Further, for objects whose observed characteristics differ purely because of viewing angle or internal structure \citep{Antonucci, Elvis}, there should be no particular difference in local environment.  This, of course, would only be the case if that structure is not correlated with fueling, as could occur, for instance, if more luminous quasars had strong winds.  It is therefore important to understand the relationship between the physical properties of AGN and their local environment, which will in turn provide insight into what aspects of AGN properties are explained by formation history, fueling, or simply by structure and orientation.  
    
    In this paper, we address this challenge by studying the nature of AGN environments.  We improve upon the most recent SDSS study, \citet{Serber}, in several ways.  We use larger samples of background photometric galaxies, as well as larger, more recent samples of spectroscopic AGN.  
    Our spectroscopic data is divided into four target samples:  Type I and Type II quasars (e.g., AGN with the highest intrinsic luminosity) and lower-luminosity Type I and Type II AGN.  The spectra of Type I AGN and quasars are characterized by broad emission lines \citep[FWHM $> 1000~\rm{km~s^{-1}}$; e.g.,][]{Haoa, Schneider}, while the spectra of Type II AGN and quasars exhibit narrow emission lines \citep[e.g.,][]{Haoa, Zakamska}.  
    We compare the overdensity of Type I quasars to the overdensity of Type II quasars as well as to lower luminosity Type I and Type II AGN.  
    Additionally, we include cuts in photometric redshift space around spectroscopic targets and the random positions to which they are compared to minimize interloping foreground or background objects, as well as marginalize over any redshift evolution of the photometric galaxy sample.  By using photometric redshift cuts, we obtain more realistic overdensity estimates and errors, and we are able to extend the study of AGN environments in the SDSS to \emph{z}~$\leqslant$~0.6.  
    
        The data samples used in this paper are discussed in Section~\ref{datasection}, and the technique for calculating overdensities is described in Section~\ref{techniquesection}.  We give a detailed discussion of the relationship of type, redshift and luminosity with the local environment of AGN in Section~\ref{discussionsection}.  Throughout the paper, we assume a concordance cosmology $\Omega_{M} = 0.3$, $\Omega_{\Lambda} = 0.7$, $H_{0} = 70$ km s$^{-1}$ Mpc$^{-1}$ (with $h = 0.7$) in order to compare to results from previous studies.  

\section{DATA}\label{datasection}
\begin{figure}
\plotone{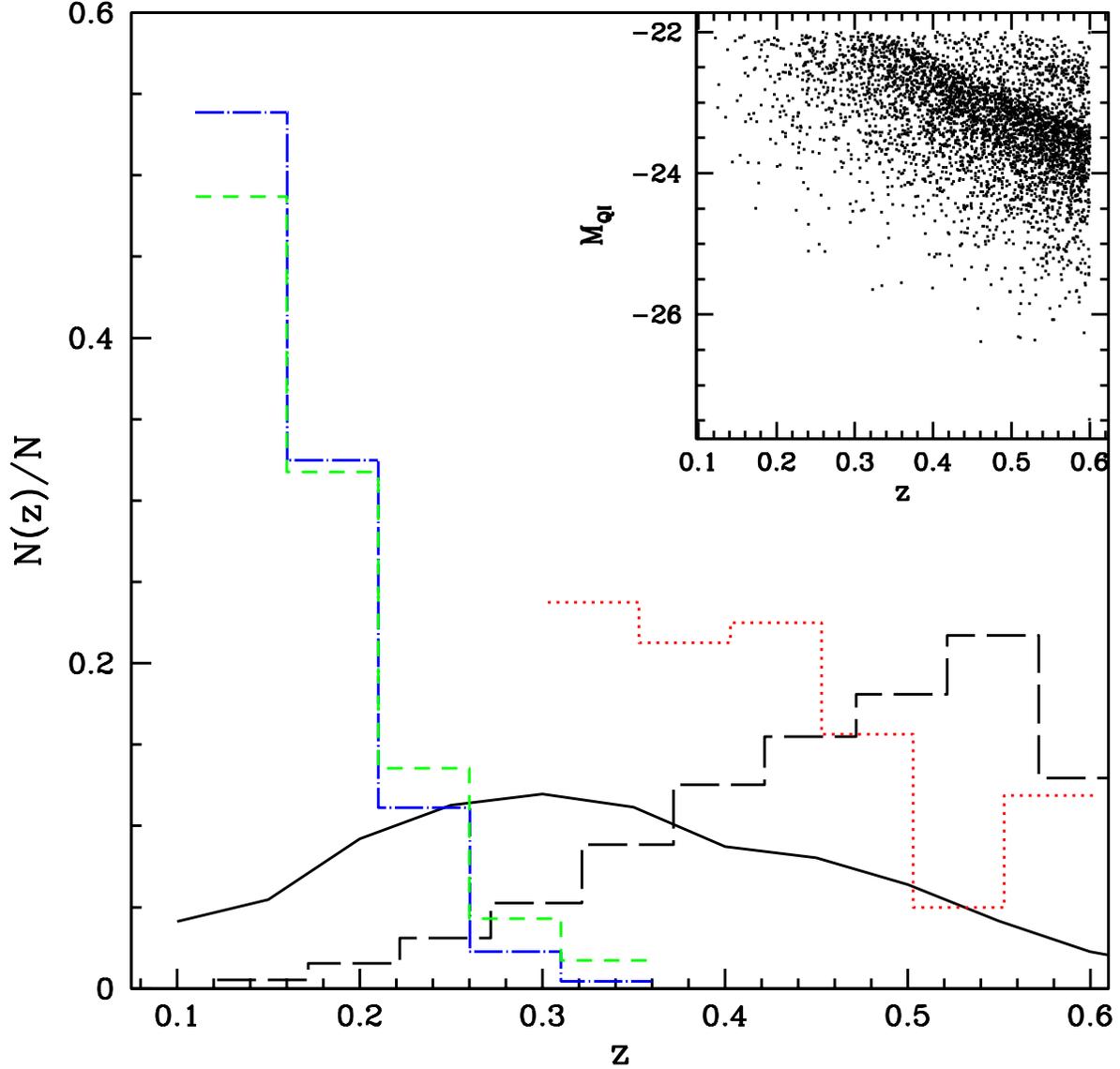}
\caption{Redshift distribution of spectroscopic targets and photometric galaxies.  The photometric galaxy redshift distribution is shown by the solid black curve.  The Type I ($-27.5 \leqslant M_{i} \leqslant -22.0$) point-source quasar distribution (4,234 objects) is shown with the black long-dashed histogram, and the Type II quasar distribution (160 objects) with the red dotted histogram.  The distribution of lower-luminosity Type I AGN (1,745) is shown with the green short dashed histogram, and the Type II AGN distribution (classified by Kewley's criteria; 3,408) is shown with the blue dot-dashed histogram.  
The inset shows absolute $i$-magnitude of Type I quasars vs. redshift.  Evidence for the $i\leqslant 19.1$ ($z \lesssim 3.0$) limit used in the quasar selection algorithm \citep{Schneider} is seen as a thicker diagonal band.  
\label{histogram_Nofz}}
\end{figure}

We study the environments of spectroscopic AGN targets by counting photometric galaxies within a 2.0$\Mpchseventy$ projected comoving distance of the target center (i.e., we consider a conical slice around the target rather than a spherical volume).  The samples of Type I quasars, Type II quasars, and lower luminosity AGN, and photometric galaxies selected from the SDSS are described below.  We apply masks to the spectroscopic samples to eliminate objects within 2.0$\Mpchseventy$ of the survey edge or an area masked out by SDSS or with seeing $>1.5\arcsec$ or $r$-band reddening $A_{r} \geqslant 0.2$.  Figure~\ref{histogram_Nofz} shows the redshift distribution of the spectroscopic and photometric samples used in our analysis. 

\subsection{Spectroscopic Targets}\label{targetsection}
Our sample of spectroscopic Type I quasar targets is drawn from the the SDSS Fifth Data Release \citep[DR5; ][]{DR5paper} Quasar Catalog \citep{Schneider}, which includes $K$-corrected absolute $i$-band magnitudes for each object.  Our quasar sample has $-27.5 \leqslant M_{i} \leqslant -22.0$, and our masking reduces the sample to $6,034$ quasar targets with $0.11 \leqslant z \leqslant 0.6$.  
In addition to absolute magnitude information, the catalog provides a flag to distinguish between resolved and point source objects.  The measured luminosity of resolved sources will likely be contaminated by starlight from the host galaxy, so we have excluded the $1,800$ quasars with extended morphology.  The final Type I quasar sample contains $4,234$ objects.  

We draw Type II quasar targets from the sample presented by \citet{Zakamska}.  After cutting the sample to match the high redshift limit of the main quasar sample ($z \leqslant 0.6$) and masking this sample as described above, we have $160$ targets with $0.3 \leqslant z \leqslant 0.6$.  

We compare the quasar targets to lower-luminosity Type I and Type II AGN from Hao (private communication) selected from SDSS DR5 spectroscopic galaxies according to the criteria laid out in \citet{Haoa}.  The classification of these galaxies as AGN depends on the strengths of the [O III] and H$\beta$ lines \citep{Haoa, Kauffmann}; therefore our low-redshift limit is set to $z = 0.11$, as this is where the [O III] ($\lambda$$\lambda$4959, 5007) lines enter the $r$-band \citep{Kauffmann}, resulting in a more uniform classification.  After masking as above, there are $1,744$ Type I AGN and $3,407$ Type II AGN following the criteria of \citet{Kewley}.  While we could have adopted the less stringent criteria of \citet{Kauffmann}, we wish to be conservative in our sample selection in this analysis and minimize the contribution of non-accretion luminosity \citep{Haoa}.  The lower-luminosity AGN samples have a redshift range of $0.11 \leqslant z \leqslant 0.33$, with the majority of sources at $z < 0.15$.  

\subsection{Photometric Galaxies}\label{photogalsection}
The photometric galaxies are drawn from the SDSS DR5 database by selecting all primary objects photometrically classified as galaxies with $r$-band extinction corrected magnitude in the range $14.0 \leqslant r \leqslant 21.0$\footnote{Note that \citet{Serber} used an $r$-band limit on their photometric galaxies rather than $i$-band as stated in their paper (W. Serber and R. Scranton, private communication).}.  All of these objects have been assigned photometric redshifts via a template-fitting technique as described in \citet{Csabai}.  Our final photometric galaxy sample of over 28 million ($28,856,324$) objects consists of only those objects that pass the flag requirements for a clean galaxy sample\footnote{as defined by \url{http://cas.sdss.org/astro/en/help/docs/realquery.asp\#flags}}.  
While we do not make explicit redshift cuts on the photometric galaxy sample, our technique (as described in Section~\ref{deltazcutsection}) effectively limits the sample to $0.06 \leqslant z \leqslant 0.65$.  

\section{TECHNIQUE}\label{techniquesection}
We count the number of photometric galaxies within a comoving radius of $2.0 \Mpchseventy$ of each spectroscopic target (e.g., spectroscopic quasar, AGN, or spectroscopic galaxy), excluding any galaxies that are within $25 \kpchseventy$ of the target.  At $z < 0.4$, $25 \kpchseventy$ corresponds to an angular size of $> 3.3\arcsec$, which is approximately twice the average seeing in DR5 \citep{DR5paper}.  At angular scales smaller than this, deblending complicates the reliable detection and measurement of faint galaxies.  

We generate a large number of random positions in the DR5 footprint area for each redshift increment of $0.001$ in our redshift range of $0.11 \leqslant z \leqslant 0.6$.  We mask the positions in the same manner as we masked the spectroscopic targets, leaving at least $1,000$ random positions that are more than $2.0 \Mpchseventy$ away from the survey edge or a masked area for each redshift value.  We note that this approach helps counteract any bias due to evolution in the photometric galaxy sample (see also Section~\ref{deltazcutsection}).  We count the number of photometric galaxies within a designated comoving distance around random positions and calculate the mean cumulative number of counts for that redshift increment as

\begin{equation}\label{randomeqn}
R_{i} = \frac{\sum_{z}R_{z}}{N_{z}}
\end{equation}
where $N_{z}$ is the number of random postions $R_{z}$ at a given redshift increment $z$.  We calculate the error corresponding to the mean random counts as
\begin{equation}\label{randomerreqn}
e_{R_{i}}^{2} = \frac{N_{z}}{N_{z}-1}(\overline{R_{i}^{2}}-\overline{R_{i}}^{2}) = \sigma_{R_i}^2
\end{equation}
which is the variance on the mean random counts at a given redshift increment $z$.  

The cumulative bincounts \emph{C$_{i}$} around spectroscopic targets are matched with the mean cumulative random bincounts $R_{i}$ (and error $e_{R_{i}}$) at the redshift increment closest to the target's redshift.  We calculate a mean overdensity $\delta_{bin}$ in a particular scale, redshift, or absolute magnitude bin as
\begin{equation}\label{avgoverdensityeqn}
\delta_{bin} = \frac{\frac{1}{N}\sum_{i}^{N}C_i}{\frac{1}{N}\sum_{i}^{N}R_i} - 1 = \frac{\overline{C_{bin}}}{\overline{R_{bin}}} - 1 = \frac{C_{bin}}{R_{bin}} - 1 
\end{equation}
where \emph{C$_{i}$} is the counts around each target in the bin, \emph{R$_{i}$} is the mean counts around random positions at the corresponding redshift, and there are \emph{N} total targets in the bin.  The error on the overdensity is determined via error propagation:
\begin{equation}\label{overdensityerreqn}
e_{\delta_{bin}}^2 = \frac{e_{C_{bin}}^2}{R_{bin}^2} + \frac{C_{bin}^2}{R_{bin}^4}e_{R_{bin}}^2
\end{equation}
where $e_{C_{bin}}$ = $\sqrt{C_{bin}}$ and $e_{R_{bin}}^2$ = $\sum_{i}^{N}e_{R_{i}}^2$.  

We will refer to the quantity of $\frac{C_{bin}}{R_{bin}}$ as the \emph{mean density}; this quantity is used to compare our results to those of \citet{Serber}.  In order to compare the environments of our various spectroscopic populations, we use the \emph{mean overdensity}, $\frac{C_{bin}}{R_{bin}}-1$, which is associated with the underlying dark matter distribution and can be more directly related to correlation analyses of clustering \citep[e.g.,][]{Padmanabhan}.  

\subsection{$\delta$\emph{z} Cut}\label{deltazcutsection}
One of the difficulties in using photometric galaxy samples for overdensity measurements is the issue of projection effects, where foreground or background objects contaminate a measurement.  We use photometric redshifts assigned to the photometric galaxies to minimize this complication.  We apply a photometric redshift cut on the galaxies so that only those galaxies which satisfy $|z_{target} - z_{photogal}| \leqslant \delta z$ are counted in each bin.  
Crucially, \emph{the same $\delta z$ cut is applied to both the spectroscopic targets and the random positions to which they are compared}, which, as noted above, are also placed at the same redshift as the spectroscopic targets.  We therefore minimize contamination by most galaxies outside of the $\delta z$ interval.  Additionally, by calculating our spectroscopic-photometric and random-photometric counts in the same $z\pm\delta z$ bin, we marginalize redshift evolution in the photometric galaxy sample outside of that $z\pm\delta z$ bin.  We assume for the purposes of this work that there is no redshift evolution in the photometric galaxy sample over this small $\delta z$ interval.  We have not accounted for the changes in photometric redshift accuracy as a function of magnitude and redshift, which we assume are negligible for this analysis.  

We verify that the projection effect issue is mitigated by the $\delta z$ cut without introducing systematics by calculating overdensities for random positions with the same redshift distribution as the Type I quasar sample.  We find that the overdensities of photometric galaxies around random positions is consistent with zero on all scales with and without the photometric redshift cut.  We use the value $\delta z=0.05$, which is large enough to encompass the effective rms error of the photometric redshifts \citep[$\Delta z_{rms} = 0.04$ for $r < 18$;][]{Budavari}, for all further analysis.  We have tested other values of $\delta z$ and find that they give consistent results, albeit with larger uncertainties for narrower cuts, which is consistent with the expectations of Poissonian sampling.  

To compare directly to the results of \citet{Serber}, who did not apply any such redshift cut, we calculate the mean density of photometric galaxies around Type I quasars with $-24.2 \leqslant M_{i} \leqslant -22.0$ and $0.08 \leqslant z \leqslant 0.4$.  At a scale of $250 \kpchseventy$, the density of photometric galaxies around quasars is $1.41 \pm 0.033$ and around $L*$ galaxies is $1.15 \pm 0.005$ without the $\delta z$ cut.  However, applying the $\delta z$ decreases the random background noise, so we measure an environment density of $2.11 \pm 0.096$ around quasars and $1.74 \pm 0.020$ around $L*$ galaxies at the same scale.  In order to confirm that we have not added systematics by using the $\delta z$ cut, we compare the relative densities of quasar environments to $L*$ galaxy environments.  The relative density of photometric galaxies around quasars to that around $L*$ galaxies is $1.22 \pm 0.029$ without the $\delta z$ cut.  The relative density does not appreciably change when the $\delta z$ cut is used: we find the relative density to be $1.22 \pm 0.057$.  
The true physical effect of the $\delta z$ cut is shown in the comparison of mean $over$densities.  At the same scale of $250 \kpchseventy$, the relative overdensity around quasars compared to around $L*$ galaxies is $2.67\pm 0.236$ without the $\delta z$ cut, but is $1.51 \pm 0.137$ with the $\delta z$ cut.  Because we have removed projection effects, the relative overdensities are lower when the $\delta z$ cut is used; however, the errors on the mean densities with the $\delta z$ cut have increased.  We believe these larger errors are more physically relevant: with no $\delta z$ cut, objects not actually correlated with the target will reduce Poissonian error estimates.  We use the $\delta z$ cut to extend our redshift range to include spectroscopic targets with $0.11 \leqslant z \leqslant 0.6$ without concern that foreground objects will contaminate the overdensity measurements.  Therefore, all subsequent analysis and figures include the $\delta z = 0.05$ cut.  

\subsection{Measurement}

\begin{figure}
\plotone{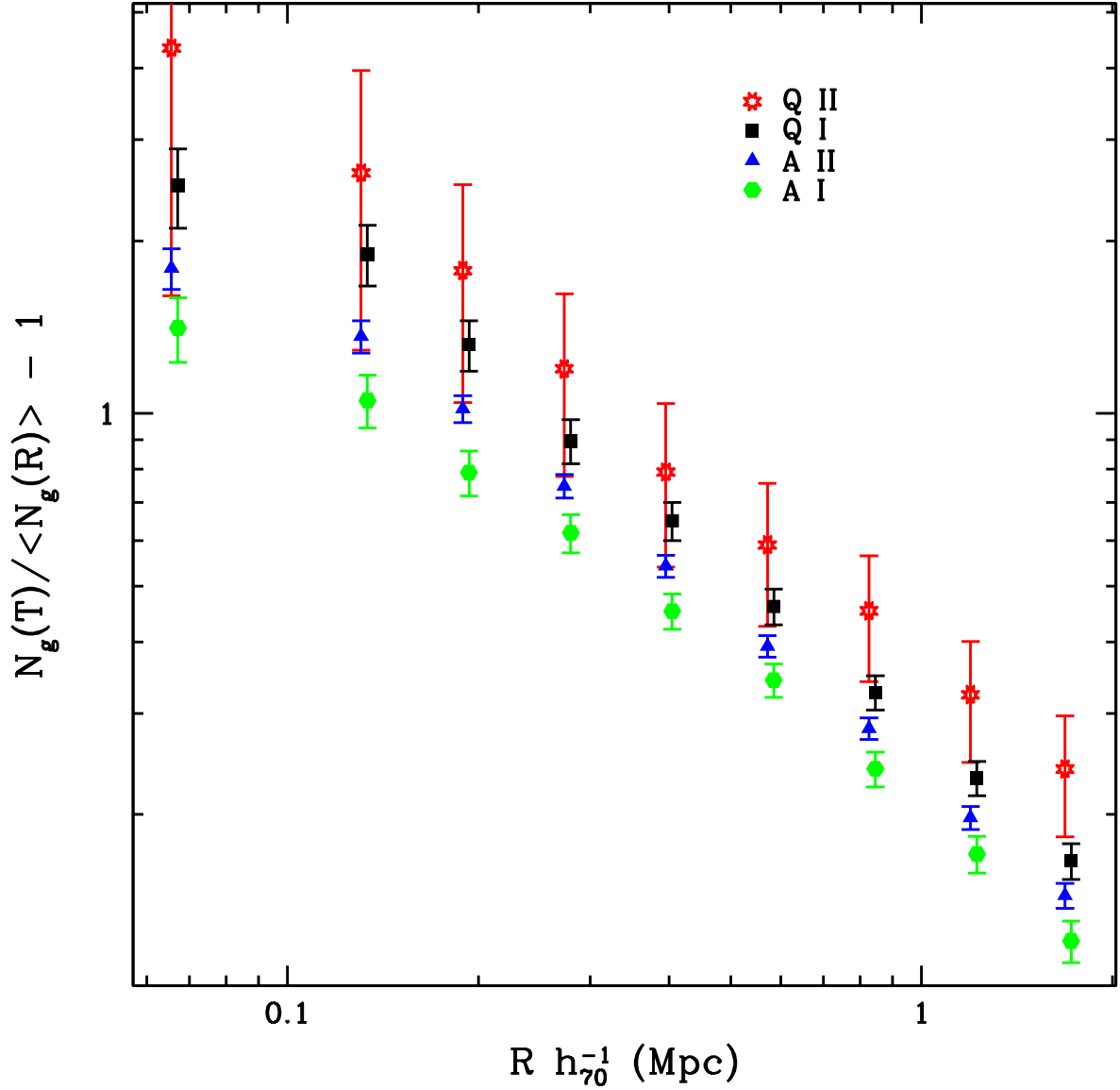}
\caption{Mean cumulative overdensity of photometric galaxies as a function of comoving scale around spectroscopic targets.  Solid black squares represent Type I quasars, open red starred points represent Type II quasars, solid blue triangles represent Type II AGN, and solid green hexagons represent Type I AGN.   Points have been offset slightly for clarity.  
\label{scale_spectargs}}
\end{figure}

In Figure~\ref{scale_spectargs}, we present the mean cumulative overdensity of photometric galaxies as a function of scale for each spectroscopic target sample.  There are clear differences: 
Quasars are in the most overdense environments at all scales, and Type II objects are in more overdense environments than Type I objects for both higher luminosity AGN (i.e. quasars) and lower luminosity AGN.  Within a scale of $\approx150\kpchseventy$, Type II quasars have an environment 1.4 times more overdense than that of Type I quasars, albeit with large errors, while the Type II AGN have an environment 1.3 times more overdense than Type I AGN.  At the same scale, the Type I quasars have environments more overdense than Type I AGN by a factor of 1.8, and Type II quasars have environments 1.9 times more overdense than Type II AGN.  Moving out to the scale of $\approx 1\Mpchseventy$, the Type II quasars again have an environment 1.4 more overdense than the environment of Type I quasars, and the Type I quasars are in environments 1.4 times more overdense than Type I AGN.  

The differences between the target samples' environment overdensities could be an effect of AGN type, however, the intertwined effects of AGN luminosity and redshift will certainly play into these differences.  In the next section, we explore how AGN type, redshift and luminosity influence the measured differences in AGN environments. 
  
\section{DISCUSSION}\label{discussionsection}
AGN luminosity and type should be key factors dictating variations in environment overdensity according to the hypothesis that information about the fueling mechanisms of different AGN will be evidenced in the AGN environments \citep[e.g.,][]{Hopkins2007}.  Additionally, the differences in overdensity between the target types seen in Figure~\ref{scale_spectargs} could be due to redshift effects, since it is obvious from Figure~\ref{histogram_Nofz} that the redshift distributions of the samples are not the same.  Therefore, in this section we investigate how the measurement of local environment varies with redshift, luminosity, and AGN type.  We first isolate each of these variables to study its impact on measured overdensities while the other two variables are held constant.  Because the variables are not independent, we consider them together in Section~\ref{allsubsec}.  

\subsection{Redshift}\label{redshiftsubsec}
\begin{figure}
\plotone{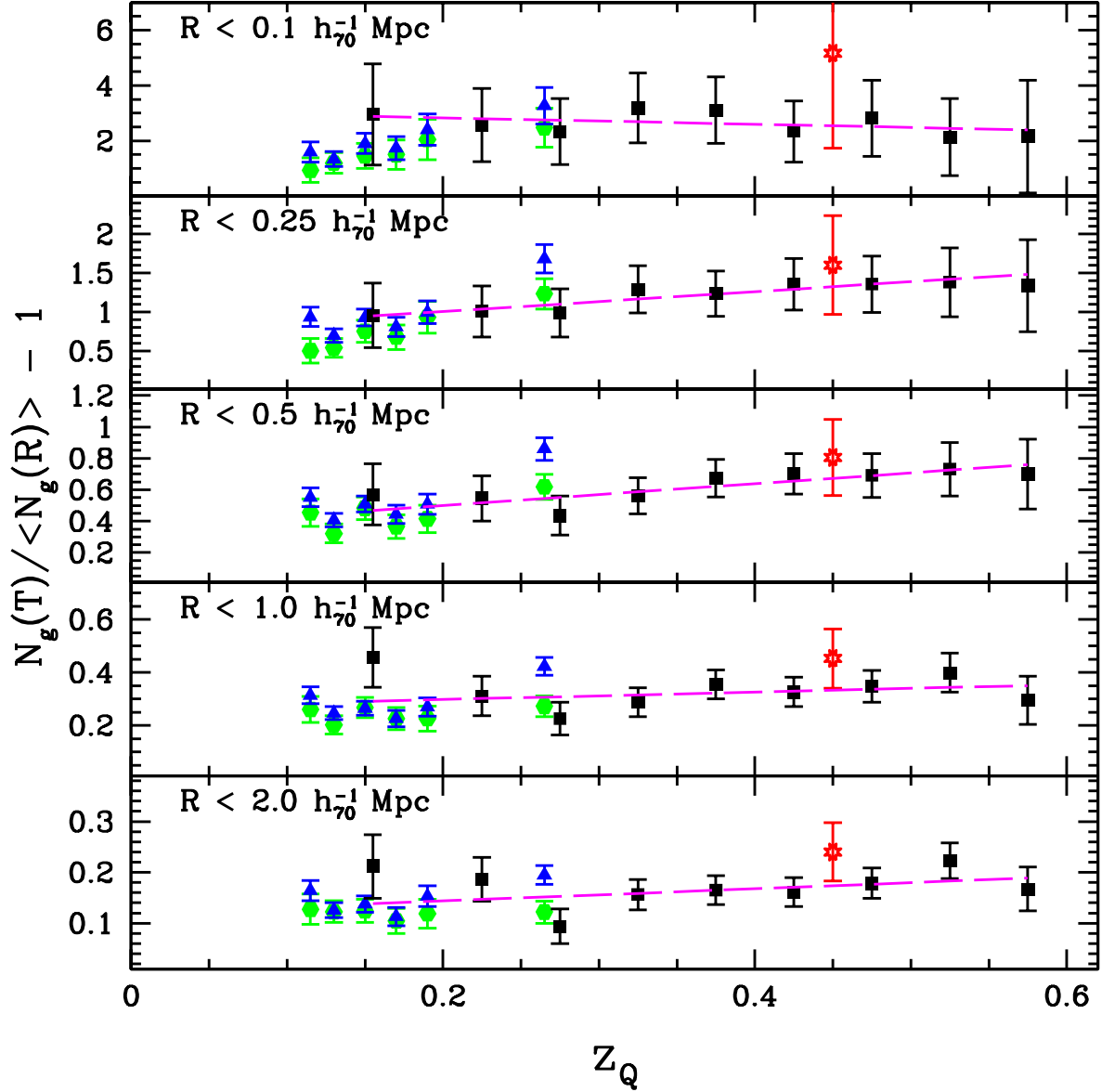}
\caption{Mean cumulative overdensity vs. redshift for spectroscopic targets.  Symbols correspond to those used in Figure~\ref{scale_spectargs}.  The magenta dashed lines are linear weighted least-squares fits for the Type I quasar sample; the parameters for these lines are given in Table~\ref{table_compareFit}.  
\label{redshift_spectargs}}
\end{figure}

We first investigate only the effects of redshift.  Figure~\ref{redshift_spectargs} shows the mean cumulative overdensity as a function of redshift for the spectroscopic target samples, which provides marginal evidence for redshift evolution in the environment overdensity of  Type I quasars.  The magenta dashed line shows the linear weighted least-squares fit to the Type I quasar environment overdensity data with redshift at different maximum radii;  the fitting parameters for these lines is given in Table~\ref{table_compareFit}. The rightmost column of the table gives the $\chi^2$ probability for each fit using the relevant degrees of freedom.  The best linear fit for the smallest scale of $R \leqslant 0.1\Mpchseventy$ has a slope of $-1.171$, which suggests decreasing overdensity with increasing redshift, and for $R \leqslant 0.25 \Mpchseventy$, the overdensity increases with redshift: the best linear fit has a slope of $1.276$.  While these fits indicate a slight redshift dependence, we also try a zero-slope linear fit (which necessarily has one less degree of freedom than the best-fit line) and find that the zero-slope fit, i.e., no redshift dependence, is also a good fit to the data and in some cases is slightly more likely.  Higher precision and higher redshift measurements will be necessary to place strong constraints on the functional form of environment overdensity evolution with redshift.  
If the environment overdensity is indeed independent of redshift, this implies that the significant differences in environment seen in Figure~\ref{scale_spectargs} are caused primarily by luminosity and type effects, rather than the influence of redshift evolution.

\begin{figure}
\plotone{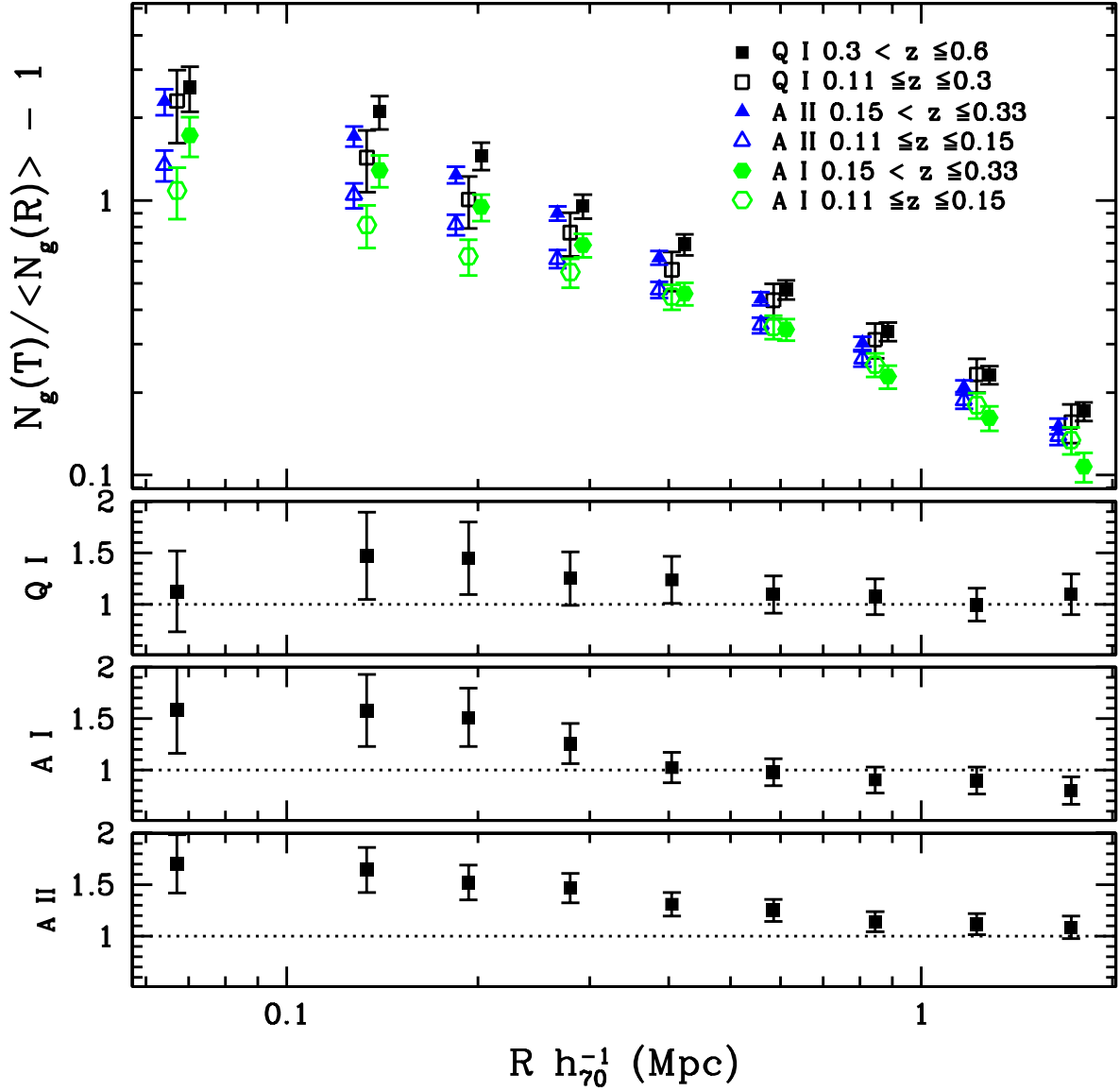}
\caption{Upper panel:  Mean cumulative overdensity of photometric galaxies around quasars and lower-luminosity AGN as a function of scale and redshift.  Points have been offset slightly for clarity.  Top lower panel:  Ratio of environment overdensity of higher-redshift Type I quasars to that of lower-redshift Type I quasars.  Middle lower panel:   Ratio of environment overdensity of higher-redshift Type I AGN to that of lower-redshift Type I AGN.  Bottom lower panel:  Ratio of environment overdensity of higher-redshift Type II AGN to that of lower-redshift Type II AGN.    
\label{scale_spectargs_z}}
\end{figure}

In the top panel of Figure~\ref{scale_spectargs_z}, we show the evolution of the mean cumulative overdensity of photometric galaxies in the environments of Type I quasar, Type I AGN and Type II AGN samples.  It is important to recall that we have placed the random points at the same redshift as the spectroscopic targets, and that we have imposed $\delta z$ cuts on the photometric galaxies (as described in Section~\ref{techniquesection}) in order to minimize the effect of redshift evolution in the photometric galaxy sample.  Therefore we can compare objects in different redshift bins.  Figure~\ref{scale_spectargs_z} demonstrates that higher redshift Type I quasars are in environments 1.24 times more overdense than the lower redshift quasars on scales $\lesssim500\kpchseventy$, while at larger scales, there appears to be little-to-no redshift evolution.  However, there is scale-dependent redshift evolution evident on scales $\lesssim1.0\Mpchseventy$ for the Type II AGN, shown in the lowest panel.  The Type I AGN begin to exhibit noticeable redshift evolution at scales $\lesssim300\kpchseventy$, where the environments of lower redshift Type I AGN are 1.26 times less dense than those of the higher redshift Type I AGN. 

We see therefore that there is some evidence for a change in local environment as a function of redshift, all else being held constant.  However, we have not yet taken AGN luminosity into account.  Even in the same redshift range, selection effects due to the magnitude-limited samples may come into play, which we investigate in Sections~\ref{lumsubsec} and \ref{allsubsec}.  

\subsection{Type}\label{typesubsec}

\begin{figure}
\plotone{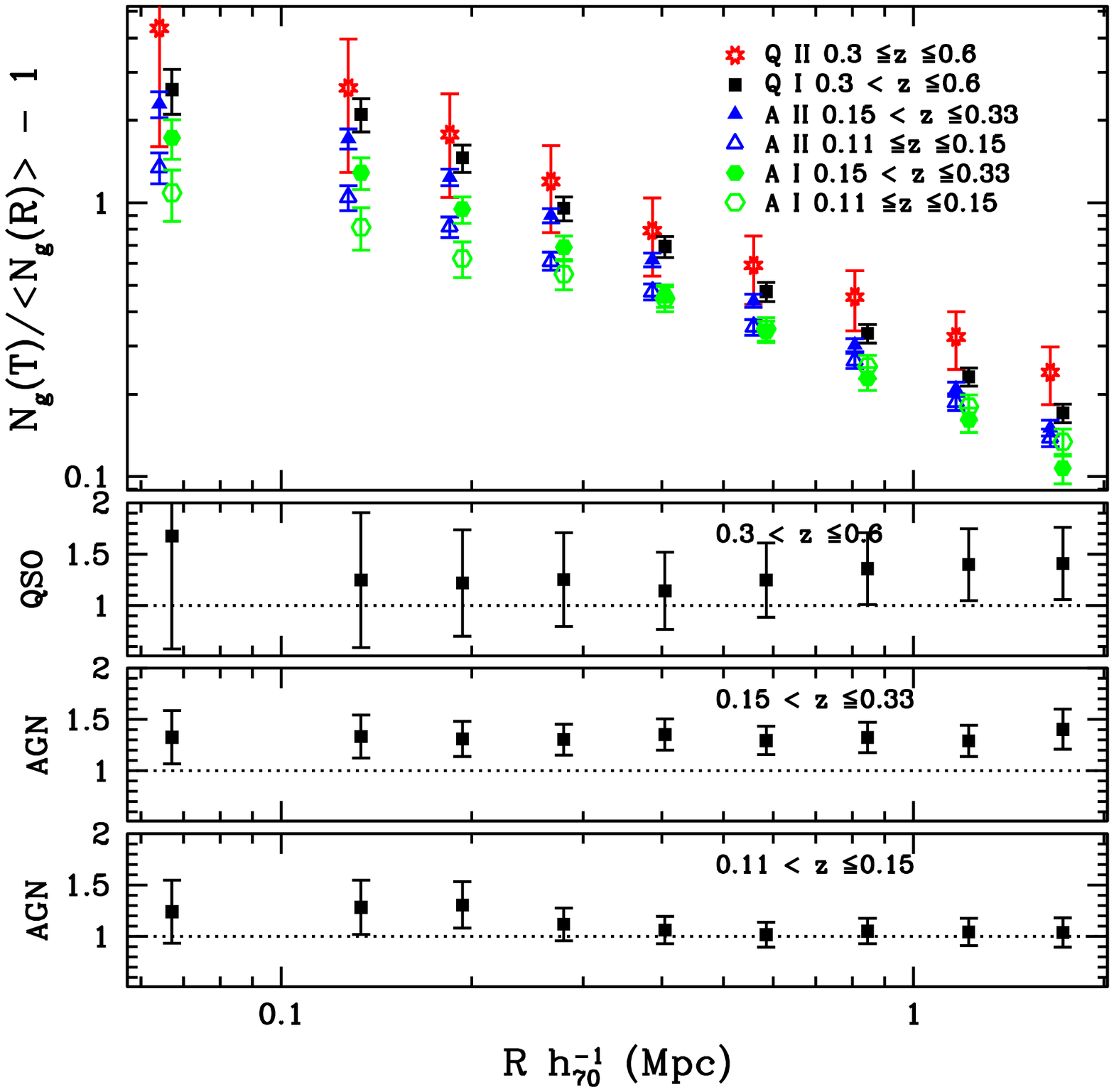}
\caption{Upper panel:  Mean cumulative overdensity of photometric galaxies around quasars and lower-luminosity AGN as a function of scale and redshift.  Points have been offset slightly for clarity. Top lower panel:  Ratio of environment overdensity of Type II quasars to that of Type I quasars in the redshift range $0.3 \leqslant z \leqslant 0.6$.  Middle lower panel:   Ratio of environment overdensity of Type II AGN to that of Type I AGN in the redshift range $0.15 < z \leqslant 0.33$.  Bottom lower panel:  Ratio of environment overdensity of Type II AGN to that of Type I AGN in the redshift range $0.11 < z \leqslant 0.15$.   
\label{scale_spectargs_type}}
\end{figure}

In Figure~\ref{scale_spectargs_type}, we identify three redshift ranges where there is overlap between our AGN samples and explore whether differences in type are reflected in the relative overdensity.  The top panel shows the overdensity as a function of scale for both types of higher-luminosity AGN (i.e. quasars) in the range $0.3 \leqslant z \leqslant 0.6$, and for both types of lower-luminosity AGN in two redshift ranges, $0.11 \leqslant z \leqslant 0.15$ and $0.15 < z \leqslant 0.33$.  The dividing redshift value of $z=0.15$ is chosen to roughly equalize the number of lower-luminosity AGN in each redshift range.  The lower three panels show the ratio of Type II environment overdensity to Type I environment overdensity in the three redshift ranges.  Again, we are able to compare objects in different redshift ranges because we have imposed $\delta z$ cuts on the photometric galaxies around both the spectroscopic targets and the random positions to which they are compared (see Section~\ref{techniquesection}) in order to account for any redshift evolution in the photometric galaxy sample and to minimize projection effects.  

Type II quasars have higher overdensity environments than Type I quasars with little scale dependence: at $R\approx1.0\Mpchseventy$, the overdensity of Type II quasar environments is a factor of 1.4 greater than the overdensity of Type I quasar environments, and at the smaller scale of $R\approx250\kpchseventy$, the Type II quasar environments have 1.2 times the overdensity of Type I quasar environments.   However, the large errors prevent us from drawing strong conclusions.  

Type II environment overdensity is again consistently about a factor of 1.3 higher on all scales than the Type I environment overdensity for lower luminosity AGN in the redshift range $0.15 < z \leqslant 0.33$.  In the lower redshift range of $0.11 \leqslant z \leqslant 0.15$, however, the Type II and Type I environment overdensities have a ratio consistent with unity until scales $R < 200\kpchseventy$, where the ratio increases to 1.3.  This agrees with the work of \citet{Koulouridis}, who found that Type II Seyfert galaxies were more likely to have a close neighbor than Type I Seyferts, using small samples of Seyfert 1 and 2 galaxies at very low redshifts.

Because we see increased overdensity for Type II AGN compared to Type I AGN on small scales in overlapping redshift ranges, we can conclude that in Figure~\ref{scale_spectargs}, the differences seen in environment overdensity between the AGN types are not primarily due to redshift evolution.  However, we have not ruled out the effects of AGN luminosity.  The Type I and Type II AGN samples are selected from magnitude-limited spectroscopic galaxy samples, which will be dominated by intrinsically more luminous sources at higher redshift.  The Type I AGN sample could be more affected by this magnitude limit, as the broad emission lines contribute more significantly to the overall flux in a given band and therefore the two AGN populations could have different average intrinsic luminosities.  

\subsection{Luminosity}\label{lumsubsec}

Unlike Type I quasars, which are targeted largely based on their strong nuclear luminosity \citep{Schneider}, the lower luminosity AGN we use 
were selected from objects classified as galaxies by the SDSS selection algorithms \citep{Haoa}.  The broad-band flux of these sources will be dominated by host galaxy starlight and/or flux from star formation, etc., which has little or no association with the nuclear luminosity.  Therefore, we focus on the Type I quasar sample for our luminosity analysis, as it spans the entire redshift range we study, and with this long redshift baseline we are best able to disentangle redshift and luminosity effects on environment overdensity.  

In order to verify that the observed evidence for evolution of environment overdensity is not due to the $i \leqslant 19.1$ ($z \lesssim 3.0$) limit imposed on Type I quasar selection in the SDSS \citep{Schneider}, we perform several tests in which we vary the apparent magnitude limit of the data.  We consider two quasar samples limited to $i \leqslant 18.9$ and to the $i \leqslant 19.1$ SDSS limit (see inset of Figure~\ref{histogram_Nofz}).  
The two magnitude-limited samples were each then separated into two luminosity bins.  We compared environment overdensity measurements of bright or dim quasars in each of the magnitude-limited samples and found no appreciable difference.  Additionally, no difference was observed when different absolute magnitude values were used to define the bright and dim samples.  In order to ensure that there is no difference between environments of quasars with $i > 19.1$, which were selected by the high-redshift targeting algorithm, and the rest of the apparent magnitude-selected sample, we performed similar tests comparing the environment overdensity of the entire quasar sample to that of the subset of quasars with $i\leqslant 18.9$ or $i > 19.1$.  In all cases, there was no appreciable change in the observed overdensity. 

\begin{figure}
\plotone{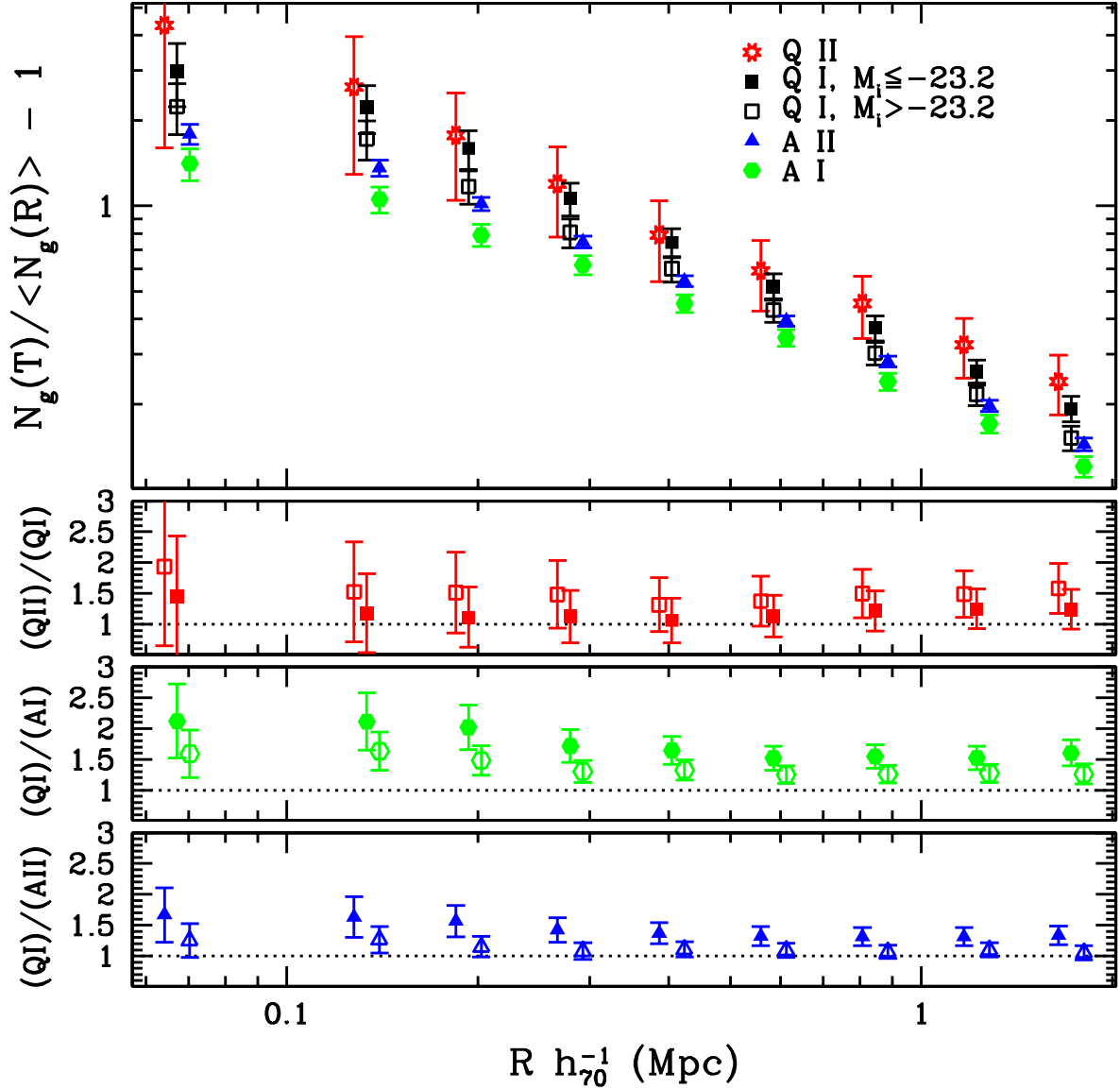}
\caption{Top panel:  Mean cumulative overdensity of photometric galaxies around Type I point-source quasars split by luminosity; low luminosity AGN; and Type II quasars.  Bright quasars have absolute magnitude $-27.5 \leqslant M_{i} \leqslant -23.2$ and dim quasars have absolute magnitude $-23.2 < M_{i} \leqslant -22.0$.  Top lower panel:  ratio of Type II quasar environment overdensity to bright (solid points) and dim (open points) Type I quasar environment overdensities.  Middle lower panel:  ratio of bright (solid points) and dim (open points) Type I quasar environment overdensities to Type I AGN overdensity.  Bottom lower panel: ratio of bright (solid points) and dim (open points) Type I quasar environment overdensities to Type II AGN.  No redshift limits have been imposed.  
\label{scale_spectargs_M}}
\end{figure}

We compare the environment overdensities of Type I quasars in two luminosity bins to the other target samples without redshift cuts in Figure~\ref{scale_spectargs_M}.  The threshold value $M_{i}=-23.2$ is chosen to give roughly equal numbers of Type I quasars in each luminosity bin: there are 2,190 (2,044) quasars with $-27.5 \leqslant M_{i} \leqslant -23.2$ ($-23.2 < M_{i} \leqslant -22.0$).  The average magnitude of the brighter (fainter) bin is $\overline{M_{i}} = -23.83$ ($\overline{M_{i}} = -22.70$).  

Type II quasars and the brighter Type I quasars are located in similarly overdense environments consistently at all scales, while the dimmer Type I quasars are located in environments slightly less overdense than the Type II quasars.  At a scale $R\approx500\kpchseventy$, the cumulative overdensity of Type II quasar environment is 1.06 times that of the brighter Type I quasars, but 1.3 times as the dimmer Type I quasars.  At the scale of $R\approx1.0\Mpchseventy$, Type II quasars have environment overdensities 1.2 times the environment overdensity of brighter Type I quasars but 1.5 times that of dimmer Type I quasars.  Again we note that the large error bars nearly overlap with unity and prevent strong conclusions.  

The more luminous Type I quasars are located in environments more overdense than Type I AGN, while there is less difference in the overdensities of dimmer Type I quasars and Type I AGN.  The environment overdensity ratio increases with decreasing scale for both brighter and dimmer Type I quasars.  At a scale $R\approx500\kpchseventy$, brighter Type I quasar environments have an overdensity 1.6 times the overdensity of Type I AGN environments with significance $\approx3\sigma$, and dimmer Type I quasar environments have an overdensity 1.3 times the overdensity of Type I AGN environments with significance $\approx2\sigma$.  At $R\approx150\kpchseventy$, the environments of brighter Type I quasars are 2.1 times as overdense ($2.4\sigma$), and the environments of dimmer Type I quasars are 1.6 times as overdense as the environments of Type I AGN ($2\sigma$).  

The ratio of Type I quasars to Type II AGN increases for both bright and dim quasars with decreasing scale, but less dramatically as the ratio to Type I AGN.  The ratio between dimmer Type I quasars and Type II AGN is approximately consistent with unity for scales $150\kpchseventy \lesssim R \leqslant 2.0\Mpchseventy$; the ratio between brighter Type I quasars and Type II AGN is 1.3 ($\gtrsim2\sigma$) for scales $R\gtrsim500\kpchseventy$.  On smaller scales, both ratios increase.  At scales $R\approx150\kpchseventy$, the ratio of brighter Type I quasars to Type II AGN is 1.6 ($\approx2\sigma$), and the ratio of dimmer Type I quasars to Type II AGN is 1.3 ($>1\sigma$).  This scale dependency could be evidence for the merger origin of quasars, since one would expect to see a higher density of environment galaxies at small scales where merger events are likely to take place \citep{Hopkins2007}. 

Evidence that dimmer quasars and lower-luminosity AGN are located in environments with similar overdensity might suggest that dimmer quasars could be a transition population between low-luminosity AGN (likely fueled in dry mergers, close encounters, or secular processes) and high-luminosity AGN (likely fueled in major mergers).  Rather than disparate populations of merger-fueled and secularly fueled AGN, there may be a continuum of galaxy interactions from major mergers to close encounters or harassment that cause AGN luminosity differences.  Alternatively, a mix of mergers and secular processes could drive the AGN population near the quasar-Seyfert divide \citep[$M_{i}\approx-22.5$;][]{Haob}.  We have compared the AGN samples without redshift cuts, but we note that in Section~\ref{redshiftsubsec} we demonstrated that evolution of quasar environments with redshift is negligible.  

The significant difference in the environments of bright Type I quasars and the environments of both Type I and Type II AGN could imply that these populations have different fueling mechanisms.  This result is consistent with results presented by \citet{Li2006, Li2008}, who find that there is only a weak link between nearby neighbors of narrow-line AGN and their nuclear activity.  

\subsection{Luminosity, Redshift, and Type}\label{allsubsec}

\begin{figure}
\plotone{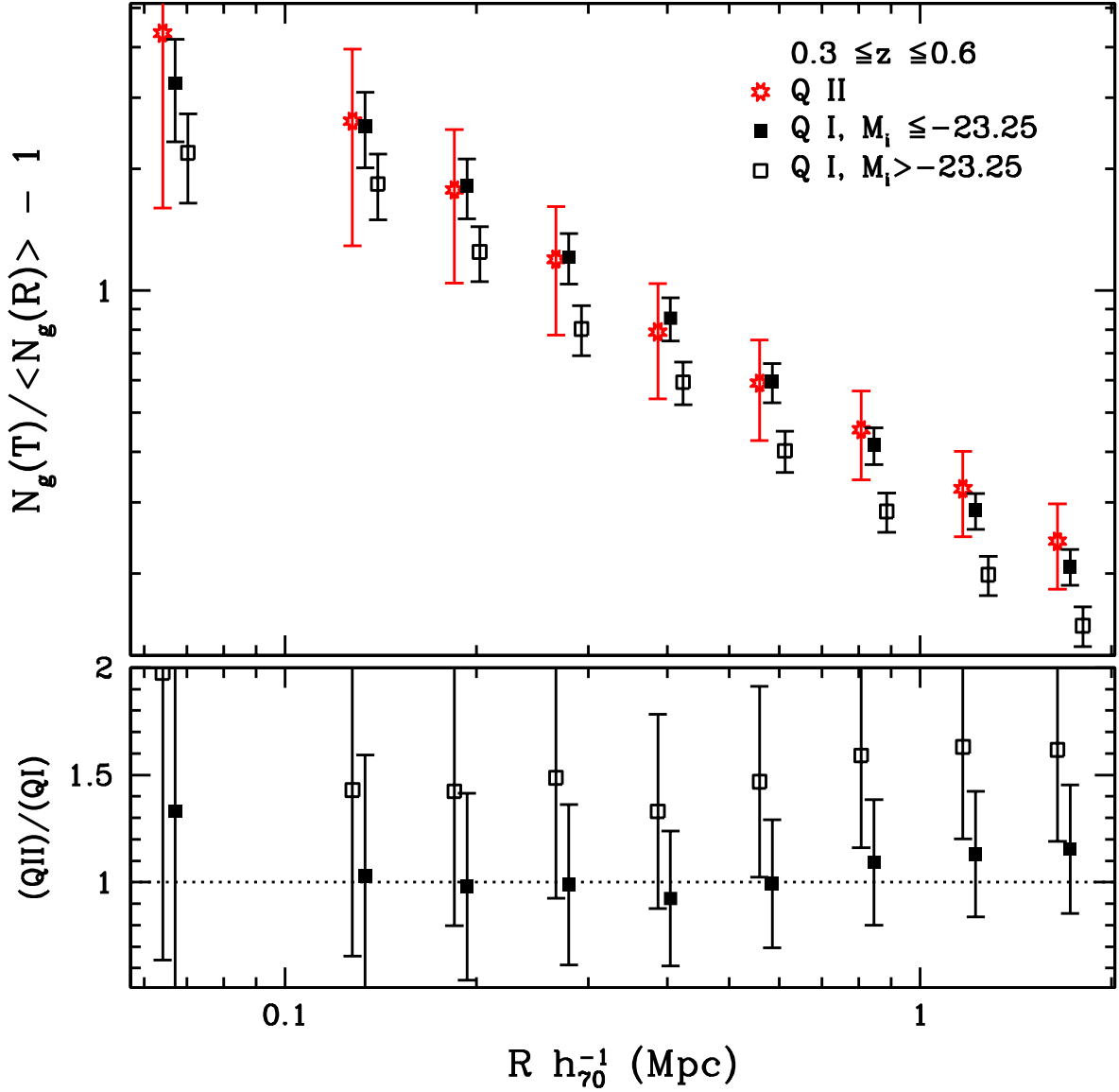}
\caption{Upper panel:  Mean cumulative overdensity of photometric galaxies around Type II quasars Type I quasars in the redshift range $0.3 \leqslant z \leqslant 0.6$.  The Type I quasars in this redshift range have been divided at $M_{i}=-23.25$ so that the luminosity bins contain approximately equal numbers of Type I quasars.  Lower panel:  Ratio of environment overdensities of Type II quasars to brighter and dimmer Type I quasars in this redshift range.  Points in both panels have been offset slightly for clarity.  
\label{scale_spectargs_Mandz_higherz}}
\end{figure}

\begin{figure}
\plotone{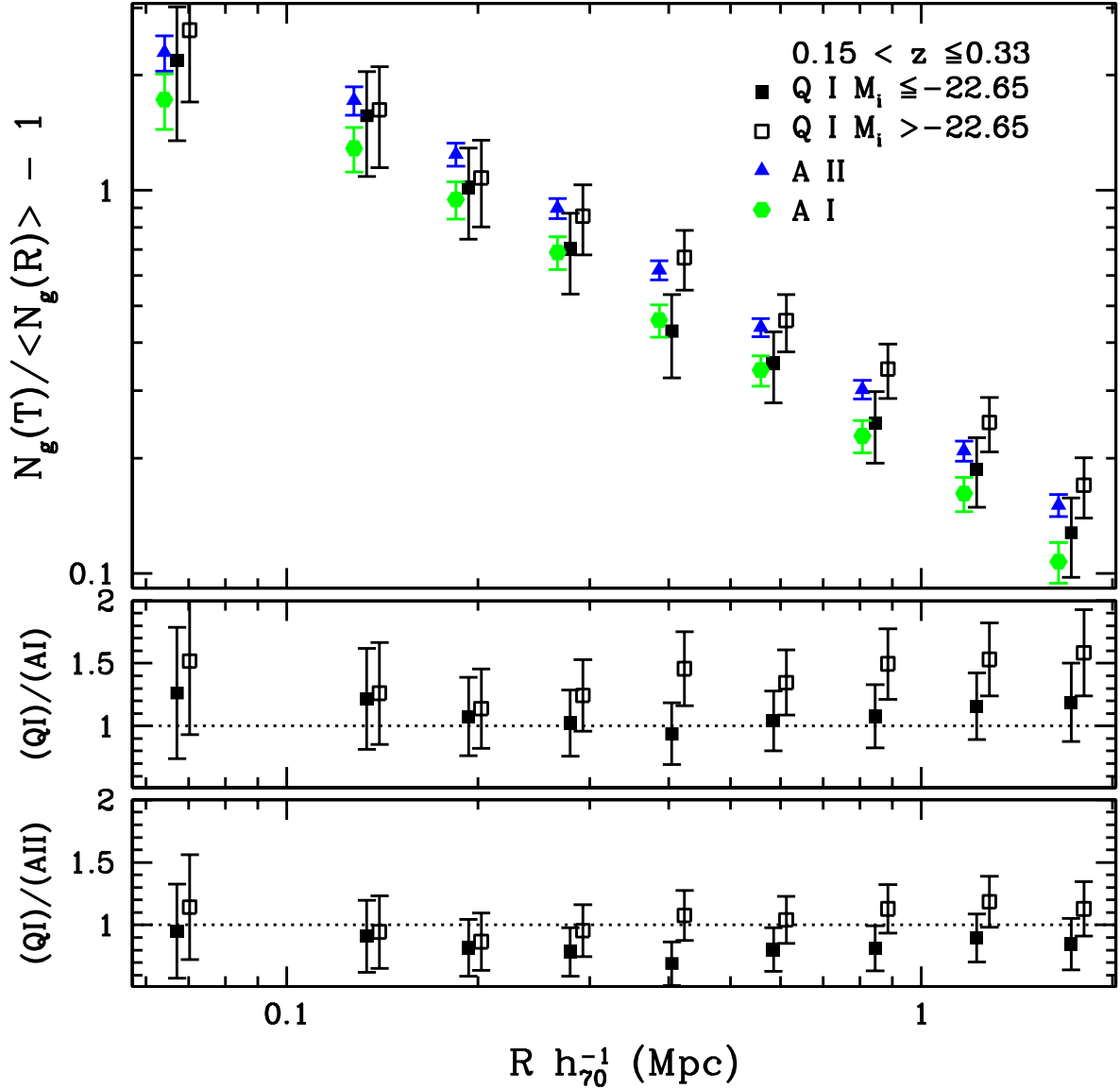}
\caption{Upper panel:  Mean cumulative overdensity of photometric galaxies around Type I quasars and Type I and II AGN in the redshift range $0.15 < z \leqslant 0.33$.  The Type I quasars in this redshift range have been divided at $M_{i}=-22.65$ so that the luminosity bins contain approximately equal numbers of Type I quasars.  Lower panels:  Ratio of environment overdensities of brighter and dimmer Type I quasars to Type I and Type II AGN in this redshift range.  Points in both panels have been offset slightly for clarity.  
\label{scale_spectargs_Mandz_lowerz}}
\end{figure}

\begin{figure}
\plotone{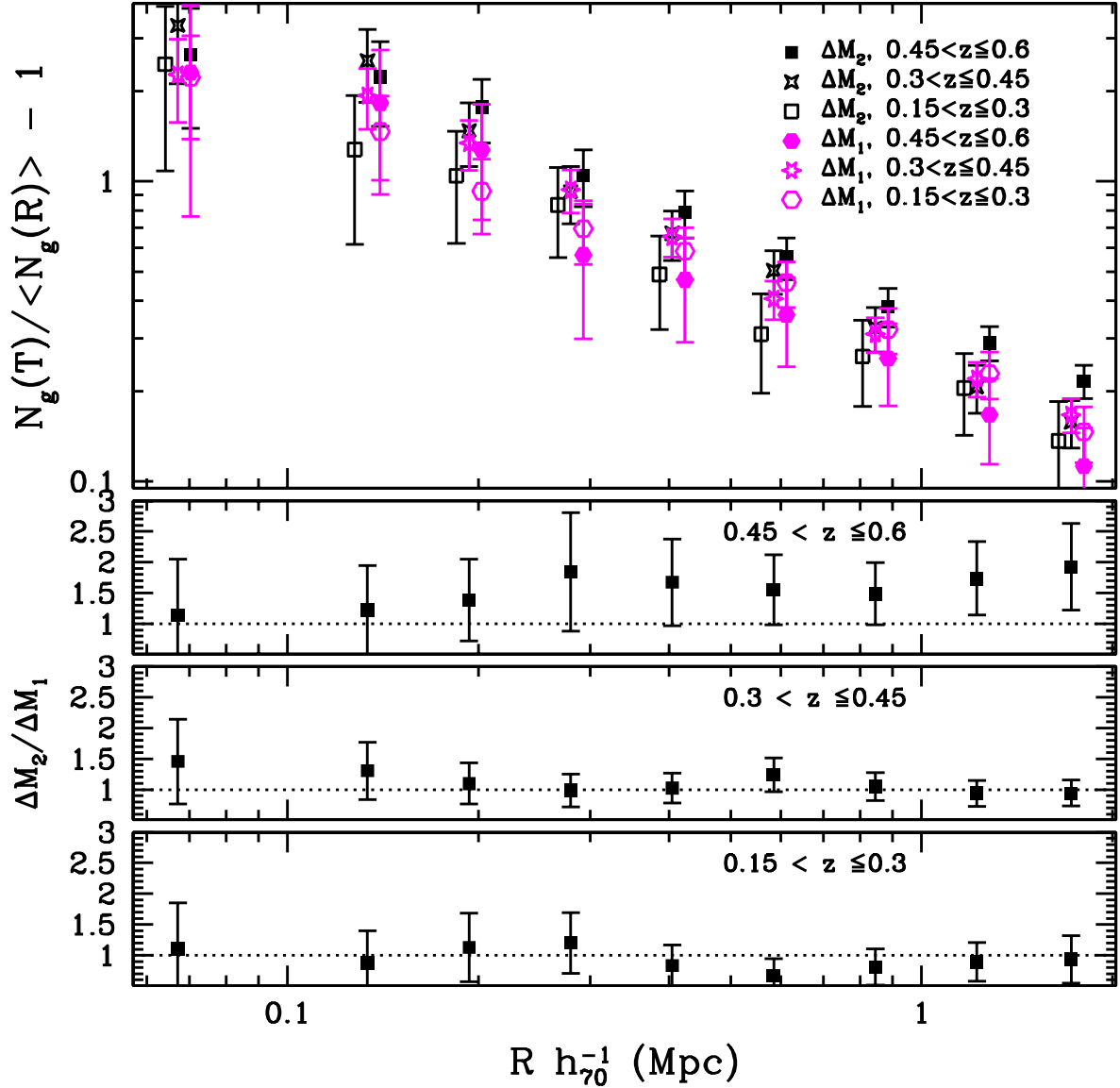}
\caption{Mean cumulative overdensity of photometric galaxies around Type I quasars with redshift for two luminosity bins, where $\Delta M_{1} : -23.0 < M_{i} \leqslant -22.0$ and $\Delta M_{2} : -24.0 < M_{i} \leqslant -23.0$.  Lower panels:  Ratio of $\Delta M_{2}$ to $\Delta M_{1}$ quasar environment overdensities in the three redshift ranges.  
\label{scale_qso_newMzcompare}}
\end{figure}

Finally, we combine our analysis of type, redshift and luminosity effects on environment overdensity in Figures~\ref{scale_spectargs_Mandz_higherz}, \ref{scale_spectargs_Mandz_lowerz}, and \ref{scale_qso_newMzcompare}.  Our $\delta z$ cuts on the photometric galaxies around the spectroscopic targets as well as around the random positions to which they are compared (as described in Section~\ref{deltazcutsection}) allow us to make meaningful comparisons of objects in different redshift ranges.  In Figure~\ref{scale_spectargs_Mandz_higherz}, Type II quasars are compared to Type I quasars in the redshift range $0.3 \leqslant z \leqslant 0.6$.  
We divide the Type I quasars into bright ($2,001$; $\overline{M_{i}}=-23.87$) and dim ($1,915$; $\overline{M_{i}}=-22.75$, about 2.8 times fainter) samples of roughly equal numbers at $M_{i}=-23.25$.  

Comparing the lower panel of Figure~\ref{scale_spectargs_Mandz_higherz} to the top ratio panel of Figure~\ref{scale_spectargs_type} shows the dramatic part luminosity plays compared to evolution alone.  The environment of Type II quasars is similar to the signature of brighter Type I quasars for all scales we measure.  The similarity of environments at small scales suggests that the differences observed between brighter Type I quasars and Type II quasars are due to a non-environmentally driven mechanism such as orientation or internal structure effects.  This in turn implies that Type II quasars are not a different cosmological population from these brighter Type I quasars.  

However, the Type II quasar environments are consistently more overdense, albeit with large errors, than those of dimmer Type I quasars on scales $R\leqslant2.0\Mpchseventy$ that we measure.  The different characteristics of the environments of the dimmer Type I quasar population from the Type II quasar population are most likely due to intrinsic luminosity differences rather than redshift differences.  
We see consistent overdensity ratios on all scales and do not see small scale effects, therefore we conclude that the difference in environment overdensity between the brighter and dimmer quasars is primarily due to mass effects.  More luminous AGN are expected to have higher mass black holes \citep[e.g.,][]{Magorrian, MarconiHunt}, which are in turn correlated with more massive dark matter halos \citep[e.g.,][]{FerrMerr, Gebhardt, Tremaine}.  
Selection effects in the magnitude-limited photometric galaxy sample could also play into the difference in overdensity between the brighter and dimmer Type I quasars.  The redshift distribution of Type I quasars in the brighter ($M_{i}\leqslant-23.25$) bin is slightly different from that of the dimmer ($M_{i}>-23.25$) bin even over the redshift range of $0.3 \leqslant z \leqslant 0.6$ (the brighter quasars have a mean redshift of 0.51, and the dimmer quasars have a mean redshift of 0.45).  The galaxies seen in the environments of brighter (higher redshift) quasars will tend themselves to be brighter, and consequently more massive, and therefore cluster more strongly than the dimmer environment galaxies \citep{Maddox, Zehavi}.  This, however, should not be a major effect.  

Figure~\ref{scale_spectargs_Mandz_lowerz} compares the environments of Type I quasars in two luminosity bins to the environments of Type I and II AGN in the redshift range $0.15 < z \leqslant 0.33$.  We use $M_{i} = -22.65$ as the threshold value for brighter and dimmer quasars in this lower redshift range to equalize the number in each luminosity bin.  The $244$ brighter ($239$ dimmer) Type I quasars have a mean magnitude of  $\overline{M_{i}}=-22.33$ ($\overline{M_{i}}=-23.30$, about 2.4 times fainter than the brighter sample).  We note that these Type I quasar samples are about a factor of seven times smaller than the Type I quasar samples in the higher redshift range, thus the measurements (and resulting interpretation) will be less precise.  

The top panel of Figure~\ref{scale_spectargs_Mandz_lowerz} shows that at for scales $R \gtrsim 300\kpchseventy$, the environments of dimmer Type I quasars are more overdense than those of brighter Type I quasars.  It appears that the situation has been reversed from Figure~\ref{scale_spectargs_Mandz_higherz}, where the environments of dimmer quasars were less overdense than the environments of brighter quasars.  However, the range of luminosity at this lower redshift range of $0.15 < z \leqslant 0.33$ is much smaller than for the higher range of $0.3\leqslant z\leqslant 0.6$.  The overall absolute magnitude distribution of these Type I quasars is skewed toward the faint end, thus the dividing value $M_{i} = -22.65$ is very close to the quasar-Seyfert divide of $M_{i} \approx -22.5$ as defined by \citet{Haob}.   Significant variation in overdensities was seen when different magnitude cuts were imposed, with the dimmer quasars consistently having higher overdensities by varying margins.  The dramatic sensitivity of results on the bright/dim dividing value emphasize that for low luminosites and redshifts, broad-band absolute magnitudes are a poor proxy for AGN luminosity.  The measured flux is more likely to be affected by galaxy starlight, star formation, etc. at this faint end.  Therefore, any attempt to use broadband magnitudes to correlate nuclear luminosity with environment will be skewed.  

With these caveats in mind, we compare the Type I quasars to lower-luminosity Type I and Type II AGN in the lower two panels of the figure.  Dimmer Type I quasar environments have overdensities greater than the Type I AGN, but the environments of brighter Type I quasars and the Type I AGN have about the same amplitude on all scales.  The lower ratio panel shows the ratio of bright and dim Type I quasars to Type II AGN.  The brigher quasars have environments with slightly lower overdensity than Type II AGN;  the environments of dimmer Type I quasars are only slightly more overdense than the environments of Type II AGN, but are consistent within the error bars.  

In Figure~\ref{scale_qso_newMzcompare} we focus on Type I quasars alone to investigate the evolution of the environment overdensity of brighter and dimmer objects.   We have chosen two luminosity intervals of one magnitude in width and compare the environment overdensities of brighter to dimmer objects in three redshift intervals. $\Delta M_{1}$ corresponds to the dimmer luminosity interval of $-23.0 < M_i \leqslant -22.0$ and contains 1,556 Type I quasars, and the brighter luminosity interval $\Delta M_2$ is $-24.0 < M_{i} \leqslant -23.0$, containing 2,036 Type I quasars.  Table~\ref{table_Fig9Details} gives the number of quasars as well as the mean magnitude in each redshift and magnitude bin.  

In the two lower redshift bins $0.15 < z \leqslant 0.3$ and $0.3 < z \leqslant 0.45$, there is little difference in the environment overdensity of brighter and dimmer quasars with little-to-no scale dependence.  
However, in the highest redshift interval of $0.45 < z \leqslant 0.6$, brighter quasars are shown to be  located in slightly more overdense environments than the dimmer quasars.  At scale of $R\approx 1.0\Mpchseventy$, the brighter quasars are located in environments with overdensity 1.5 times that of the dimmer quasars.  The brighter quasars have environments with overdensity 1.4 times the overdensity of dimmer quasar environments at a scale of $R\approx250\kpchseventy$, and then the ratio begins to drop toward unity at the innermost scales.  However, the large errors are nearly consistent with unity on all scales we measure.  

It appears, therefore, that there is again slight evidence for some redshift evolution of Type I quasar environments, but it is mainly manifested at the highest redshift range.  This emphasizes the need for additional studies of the environments around higher redshift Type I quasars.  We caution that the increased overdensity at higher redshift may be influenced by the fact that there are nearly three times the number of brighter quasars as dimmer quasars in the interval $0.45 < z \leqslant 0.6$ (see Table~\ref{table_Fig9Details}).  In the range $0.3 < z \leqslant 0.45$, the number of bright quasars is closer to the number of dim quasars, while in the range $0.15 < z \leqslant 0.3$, the dim quasars outnumber the bright quasars by more than a factor of two.  The change in overdensity ratio will also be affected by the change in mean luminosity of the dimmer quasar sample with increasing redshift.  While the bright quasar luminosity changes only by 0.07 magnitudes, the mean dim quasar luminosity changes by 0.17 magnitudes.  Therefore we cannot draw strong conclusions, but reiterate the need for higher precision and higher redshift measurements of quasar environments.   

\section{CONCLUSIONS}

Our work sheds new light on the nature of AGN environments and their relationship to the type, luminosity, and redshift of the AGN itself.  We have used larger samples of AGN targets and imposed a photometric redshift cut on the nearby photometric galaxies in order to minimize projection effects and to account for redshift evolution of the photometric galaxy sample.  By using photometric redshift cuts, we obtain more realistic overdensity estimates and errors for the local environments of AGN of various luminosities and types.  

There are two main pictures through which observed differences in AGN can be interpreted.  In the merger models presented by \citet{Hopkins2006}, differing fueling mechanisms trigger AGN activity at different luminosities.  Unified models such as those presented by \citet{Antonucci} or \citet{Elvis} ascribe observed differences in AGN to viewing angle or structure.  These pictures are not mutually exclusive; in fact, we show that the interdependency between the variables of type, luminosity and redshift play into the subtleties of distinguishing the pictures based on environment overdensity measurements.  

From our analysis, we can draw the following conclusions:
\begin{itemize}
\item Type II quasars are shown to have similar environments as brighter Type I quasars in the same redshift range on all scales that we study, which suggests the observational differences in Type I and Type II quasars are driven by orientation and/or structure and not by cosmological evolution.  

\item Lower-luminosity Type II AGN have environments consistently more overdense than lower-luminosity Type I AGN by a factor of 1.3 at all scales in the redshift range $0.15<z\leqslant0.33$, but there is scale dependence in the environment overdensity ratio of Type II to Type I AGN for redshifts $0.11\leqslant z\leqslant0.15$.  

\item There is marginal evidence for redshift evolution of Type I quasar environments on all scales, especially for $0.45 \leqslant z \leqslant 0.6$, not noted in previous studies.   However, this evolution is not the primary explanation for the environment overdensity differences seen between Type I quasars and Type II quasars, and between Type I AGN and Type II AGN.  In order to place strong constraints on the functional form of this redshift evolution,  it is necessary to acquire higher precision measurements and higher redshift measurements.  
\end{itemize}

In our current work, we have explored the effects of redshift, type and luminosity on the measurement of AGN environments in order to better understand the evolutionary nature of AGN and their general relationship with galaxies. There are several ways in which our existing work can be improved. First, when we measured the redshift effects of the AGN environments, we were confronted by the selection effects that result from our use of a magnitude-limited photometric galaxy sample. The standard way to overcome this effect is to employ volume-limited samples of galaxies, which can be done by using photometric redshifts to derive absolute magnitudes \citep[e.g.,][]{ Ross}. This approach provides a more accurate technique to compare AGN environments across different redshift ranges.
Second, as we have discussed in Section~\ref{allsubsec}, broad-band absolute magnitudes are not the best way to measure the nuclear luminosity of AGN.  Accretion luminosity, which can be more easily compared to the predictions of quasar formation models, can be better quantified by using emission line luminosities, such as O[III].  Third, our current analysis simply studies the local environment of AGN by measuring the overdensity of all galaxies near the AGN.  This approach, while useful, ignores a great deal of information.  The physical characteristics of the galaxies in the environment provide additional valuable insight into the relationship between AGN and galaxies \citep[see, e.g.][]{Barr, Coldwell2003, Coldwell2006}.  Finally, by exploring the relationship between AGN environments and intrinsic physical properties of the AGN, such as black hole mass and multiwavelength emission, we can further refine our understanding of AGN fueling mechanisms.  

\acknowledgements

We are grateful to Lei Hao for providing us with catalogs of AGN from DR5.  We thank Will Serber, Ryan Scranton, and Nikhil Padmanabhan for helpful conversations.  We also thank the referee for comments that have greatly improved the manuscript.  

We acknowledge support from Microsoft Research, the University of Illinois, and NASA through grants NNG06GH156 and NB 2006-02049. The authors made extensive use of the storage and computing facilities at the National Center for Supercomputing Applications and thank the technical staff for their assistance in enabling this work. 

Funding for the creation and distribution of the SDSS Archive has been provided by the Alfred P. Sloan Foundation, the Participating Institutions, the National Aeronautics and Space Administration, the National Science Foundation, the U.S. Department of Energy, the Japanese Monbukagakusho, and the Max Planck Society. The SDSS Web site is http://www.sdss.org/.

The SDSS is managed by the Astrophysical Research Consortium (ARC) for the Participating Institutions. The Participating Institutions are The University of Chicago, Fermilab, the Institute for Advanced Study, the Japan Participation Group, The Johns Hopkins University, the Korean Scientist Group, Los Alamos National Laboratory, the Max-Planck-Institute for Astronomy (MPIA), the Max-Planck-Institute for Astrophysics (MPA), New Mexico State University, University of Pittsburgh, University of Portsmouth, Princeton University, the United States Naval Observatory, and the University of Washington.

\clearpage


\begin{deluxetable}{c c c c c}
\tablecolumns{5}
\tablehead{
\colhead{$R_{max}$}&
\colhead{slope}&
\colhead{intercept}&
\colhead{$\chi^{2}$}&
\colhead{P($\chi^{2}$,$\nu$)}
}
\tablecaption{Linear least-squared fit parameters for quasar environment data in Figure~\ref{redshift_spectargs}.  
\label{table_compareFit}}
\startdata
$2.0\Mpchseventy$&0.119&-0.880&6.84&0.3363\\
&0.0&-0.833&8.10&0.3241\\
\cline{1-5}
$1.0\Mpchseventy$&0.140&-0.730&5.58&0.4719\\
&0.0&-0.677&6.04&0.5352\\
\cline{1-5}
$0.5\Mpchseventy$&0.691&-0.638&1.62&0.9514\\
&0.0&-0.388&4.17&0.7603\\
\cline{1-5}
$0.25\Mpchseventy$&1.276&-0.249&0.421&0.9987\\
&0.0&0.200&1.91&0.9646\\
\cline{1-5}
$0.1\Mpchseventy$&-1.171&2.066&0.628&0.9959\\
&0.0&1.635&0.716&0.9982\\
\enddata
\end{deluxetable}

\begin{deluxetable}{c c c}
\tablecolumns{3}
\tablehead{
\colhead{Redshift Range}&
\colhead{$-24.0 < M_{i} \leqslant -23.0$}&
\colhead{$-23.0 < M_{i} \leqslant -22.0$}
}
\tablecaption{Details for data used in Figure~\ref{scale_qso_newMzcompare}. \label{table_Fig9Details}}
\startdata
$0.15 < z \leqslant 0.3$& 85, $-23.39$&211, $-22.46$\\
$0.3 < z \leqslant 0.45$& 502, $-23.39$&788, $-22.57$\\
$0.45 < z \leqslant 0.6$& 1445, $-23.46$&552, $-22.63$
\enddata
\end{deluxetable}

\clearpage

\clearpage

\clearpage

\clearpage

\clearpage 

\clearpage 

\clearpage

\clearpage

\clearpage

\clearpage

\end{document}